# Coupled anisotropic weak topological states and Floquet mixed-parity altermagnetism in two-dimensional Su-Schrieffer-Heeger models

Kunyuan Feng,[1] Xibin Liu,[1] Chenchen Liu,[1] Siyuan Liu,[1] Lixiu Guan,[1] Xiaobiao Liu,[2] François M. Peeters,[3] and Linyang Li[1,*]

[1]*School of Science, Hebei University of Technology, Tianjin 300401, China*

[2]*School of Science, Henan Agricultural University, Zhengzhou 450002, China*

[3]*Department of Physics, University of Antwerp, Groenenborgerlaan 171, Antwerp B-2020, Belgium*

[*]Contact author: linyang.li@hebut.edu.cn

**ABSTRACT**

Su-Schrieffer-Heeger (SSH) topological systems and altermagnetic (AM) states are two important research areas in condensed matter physics. Realizing the coupled states between the SSH lattices and AM phanse in a single syetem remains challenging. However, the Floquet engineering change it. Our work constructs five two-dimensional-(2D-) SSH models to describe the coupled phases of SSH weak topological state and AM order by tight-binding (TB) method. The evolution of band structures, spin-splitting, and topological phase transitions under the circularly polarized light (CPL) and relative atomic displacement (RAD) in 2D SSH lattices were systematically investigated. The results reveal that the inequality of hopping parameters ($t_1$ and $t_2$) serves as the fundamental origin of SSH topological states and AM order. For the 2D nonmagnetic state, anisotropic weak topological states with Zak phase governed edge states are realized by unit cell selection, similar to the conventional 1D SSH model. The collinear antiferromagnetic state preserves the spin-degenerate band structure and intrinsic weak topological properties. Furthermore, the Floquet engineering introduces the AM phase of odd-parity p-wave while the RAD introduces the AM phase of even-parity d-wave. By combining the two effects, the mixed-parity (non-odd/non-even parity) AM phase could be realized, achieving the simultaneous control of the light field of spin-splitting and topological edge states. The physical mechanisms of Floquet engineering and RAD for the 2D rectangular SSH lattice are also from the inequality of $t_1$ and $t_2$, which can be not only fully understand by the TB methods, but also in good agreement with the first-principle calculations of 2D carbon-based materials. This work establishes an effective theoretical platform for coupling anisotropic SSH weak topological states and AM orders with multi-parities in 2D systems.

## INTRODUCTION

Topological and altermagnetic (AM) states are major areas of research in condensed matter physics, which can be regared as nontrivial properties for conventional band strucutres and antiferromagnetism (AFM)[1-12]. One of the fundamental theories for realizing topological states is the Su-Schrieffer-Heeger (SSH) model, which was developed in 1979 to describe conductive polymers[13,14]. Since its proposal, the one-dimensional-(1D-)SSH model has long been confined to theoretical analysis[1-3,14], as it has consistently lacked a reliable experimental platform. Observations of related topological edge states have only been possible through artificial simulation platforms such as cold atoms and photonic crystals[15-17]. It was not until 2024 that Nakayama et al. first experimentally observed SSH topological states in three-dimensional (3D) helical Te chains[18]. This crystal is formed by the parallel stacking of a large number of helical 1D-SSH chains and exhibits weak topological features with anisotropy. Therefore, extending the parallel stacking of 1D-SSH chains to a two-dimensional (2D) system represents a highly valuable research approach. However, most existing 2D-SSH models are built on highly symmetric square/hexagonal lattices[19-21]. Although these models retain the characteristics of SSH, they cannot achieve anisotropic topological responses similar to the experimental 3D helical Te chains, and the relevant theoretical framework still needs to be refined.

At the same time, the AM has rapidly emerged as a new frontier in magnetic states[7,11,22]. Unlike conventional AFM, AM has zero net magnetization in real space and nonrelativistic alternating spin-splitting in momentum space, providing a nontrivial platform for spintronic applications[23-30]. Based on the intrinsic symmetry classification of spin-splitting textures, AM can be divided into two groups including even-parity (*d*-, *g*-, and *i*-wave) and odd-parity (*p*- and *f*-wave) systems[31-34].

Building upon these theoretical findings, extensive efforts have been made to identify materials with AM properties[35-46]. In addition, conventional AFM materials can spontaneously generate alternate spin-splitting by various external field methods to achieve the AM order[39,45-47]. Among these, Floquet engineering is a representative approach for artificially inducing and controlling AM states[46,48]. However, current research in this area still faces two significant limitations. Firstly, most theoretical models have largely been limited to hexagonal lattices. The responses of more lattice types to Floquet engineering remain to be investigated, such as the 2D-SSH model. Secondly, the Floquet engineering is limited to introduce the AM states from the conventional AFM materials. The responses of different AM states to Floquet engineering remain to be investigated.

In this letter, starting from the 1D-SSH model, we have established the theoretical framework for five Different types of 2D-SSH theoretical models, inlcluding nonmagnetic (NM), AFM, and AM states, as shown in Table 1. The two regulation methods are used to produce the AM states, including Floquet engineering[48] (circularly polarized light, CPL) and relative atomic displacement (RAD) leading to antiferroelectric (AFE) polarization[39]. Here, the tight-binding (TB) methods, Wannier function, and the first-principle calculations were used. Beyond the theoretical models, corresponding four 2D dumbbell carbon-based materials were used to further confrim our proposed SSH models, as shown in Table 1.

**Table 1.** Five Different types of 2D-SSH theoretical models and corresponding four 2D dumbbell carbon-based materials.

| Theoretical models | 2D materials |
|---|---|
| 2D-SSH-NM | $C_4N_2/C_6N_2Si_2$ |
| 2D-SSH-AFM | $C_4N_2/C_8N_2$ |
| 2D-SSH-AM-CPL | $C_8N_2$ |
| 2D-SSH-AM-RAD | $C_{12}H_6$ |
| 2D-SSH-AM-RAD-CPL | $C_{12}H_6$ |

**SSH model of 1D and 2D systems**

We briefly review the classic SSH model and corresponding topological characteristics. Figure 1(a) shows the 1D-SSH model with a zigzag shape, where the A and B sites in a unit are coupled by strong (solid line) or weak (dashed line) hopping. The Hamiltonian is[49]

$$H = \sum_{\langle ij \rangle} \left( t_{ij} c_i^{\dagger} c_j + h.c. \right),$$

where $t_{ij}$ is the electron hopping parameter between the $i$-th and $j$-th atoms, and $c^{\dagger}_i$ and $c_j$ are the creation and annihilation operators, respectively. We only consider nearest-neighbor hopping including intracellular ($t_1$) and intercellular ($t_2$) hopping. The topological phase can be determined by the Zak phase[49-54]. For $|t_1| > |t_2|$, Zak phase is equal to 0, and the system is topologically trivial. For $|t_1| < |t_2|$, Zak phase is equal to π, the system is topologically nontrivial. The case of $|t_1| = |t_2|$ is a critical point, the band structure undergoes a bandgap closure. When Zak phase is 0, the system has no end states [Fig. 1(g)]. When Zak phase is π, there are two zero-energy modes, corresponding to the end states [Fig. 1(j)]. For the two cases, the reversal of parity at the X point corresponds to the topological phase transition.

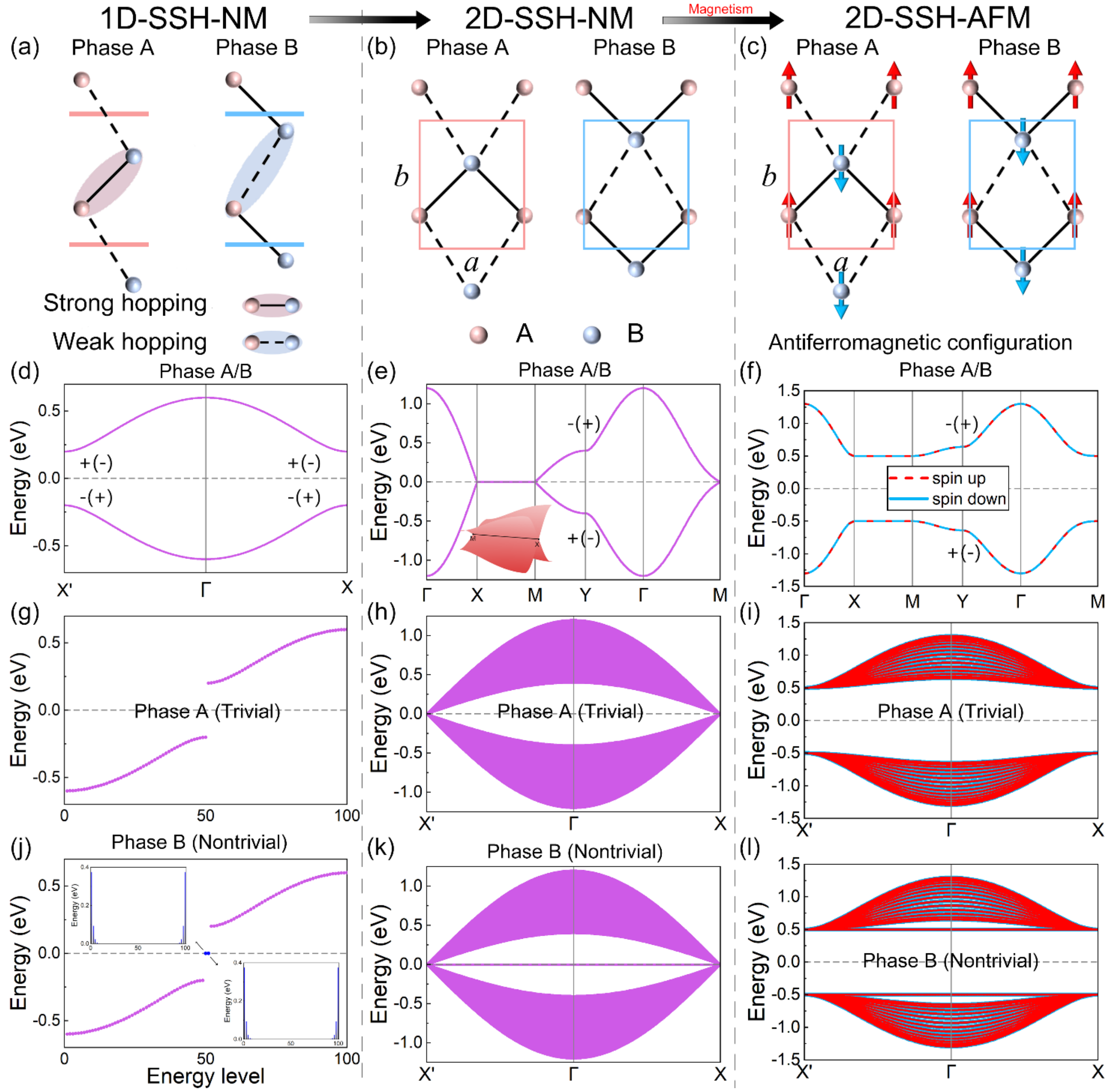

**FIG. 1.** (a)-(c) Two SSH Phase A/B for 1D-SSH-NM, 2D-SSH-NM and 2D-SSH-AFM. Red/blue balls denote A/B atoms, and solid/dashed lines mark strong/weak hopping. (d)-(f)Band structures of Phase A/B. (g)-(i)/(j)-(l) Trivial/Nontrivial edge (end) states of trivial Phase A/B.

From the 1D-SSH-NM model, the 2D-SSH-NM model with a rectangular lattice also has two distinct phases [Fig. 1(b)]. The Phase A satisfies $|t_1| > |t_2|$ and the Phase B satisfies $|t_1| < |t_2|$ and they have the same band structures [Fig. 1(e)]. However, the parity reversal at the high symmetry Y

indicates the occurrence of the topological phase transition. We calculated the edge states of the 1D nanoribbons to confirm the topological difference. The edge states along the *x*-direction cannot distinguish between the two phases, as shown in Figs. S2(a)-(c) in the Supplementary Materials (SM) Part I[55], but the edge states along the *y*-direction can effectively distinguish them [Figs. S2(d)-(e)]. The Hamiltonian of the 2D-SSH-NM model can explain these phenomena. The Hamiltonian is

$$H_{2D}=\begin{bmatrix}0 & \left(1+e^{-ik_xa}\right)\left(t_1+t_2e^{-ik_yb}\right)\\ \left(1+e^{ik_xa}\right)\left(t_1+t_2e^{ik_yb}\right) & 0\end{bmatrix}=\begin{bmatrix}0 & \rho_x\left(k_x\right)\rho_y\left(k_y\right)\\ \rho_x^*\left(k_x\right)\rho_y^*\left(k_y\right) & 0\end{bmatrix},$$

where $\rho_x\left(k_x\right)=1+e^{-ik_xa}=\left|\rho_x\left(k_x\right)\right|e^{i\phi_x\left(k_x\right)}$ and $\rho_y\left(k_y\right)=t_1+t_2e^{-ik_yb}=\left|\rho_y\left(k_y\right)\right|e^{i\phi_y\left(k_y\right)}$, with lattice constants *a* (*x*-direction) and *b* (*y*-direction). The $\phi_l(k_l)$ is the radian angle of $\rho_l(k_l)$, ranging from $-\pi\leq\phi_l(k_l)\leq\pi$. The $\phi_x\left(k_x\right)=\arctan\dfrac{-\sin\left(k_xa\right)}{1+\cos\left(k_xa\right)}$ and $\phi_y\left(k_y\right)=\arctan\dfrac{-\sin\left(k_yb\right)}{t_1/t_2+\cos\left(k_yb\right)}$, with the corresponding eigenvectors being $u\left|k\right\rangle=\dfrac{1}{\sqrt{2}}\begin{pmatrix}\pm e^{i\phi_x\left(k_x\right)}e^{i\phi_y\left(k_y\right)}\\ 1\end{pmatrix}$. The $\phi_x(k_x)$ is independent of $t_1$ and $t_2$, while the $\phi_y(k_y)$ relies on both hopping parameters. The Zak phase along path *l* is defined as $\varphi_l=-i\int_{-\pi}^{\pi}\left\langle u_n\left(k\right)\right|\frac{\partial}{\partial_{k_l}}\left|u_n\left(k\right)\right\rangle dk_l$ ($\varphi_{Zak}=\varphi_l$)[56]. For $k_x$, $\varphi_x=\frac{1}{2}\int_{-\pi/a}^{\pi/a}\frac{d\phi_x\left(k_x\right)}{dk_x}dk_x=0$ in both phases, and the nanoribbon stays trivial edge states [Figs. S2(a)-(c)]. The $\phi_x(k_x)$ remains continuous without jumps, as shown in Figs. S2(g)-(i). We analyzed the topological phase from the $k_y$ afterwards. For Phase A with $|t_1|>|t_2|$, the system is topologically trivial at $\varphi_y=\frac{1}{2}\int_{-\pi/b}^{\pi/b}\frac{d\phi_y\left(k_y\right)}{dk_y}dk_y=0$, as shown in Figs. 1(h) and S2(d). As the hopping strength gradually changes until $|t_1|=|t_2|$, the band structure evolves as shown in Fig. S1(b). The conduction and valence band structures touch along the high symmetry paths MY ($k_y=\pi$) forming a closed nodal line along the first Brillouin zone (BZ). For Phase B with $|t_1| < |t_2|$, at $\varphi_y = \frac{1}{2}\int_{-\pi/b}^{\pi/b} \frac{d\phi_y(k_y)}{dk_y} dk_y = -\pi$, the system corresponds to topologically nontrivial edge states [Fig. 1(k)]. Moreover, the $\phi_y$ ($k_y$) also changes with hopping strength [Figs. S2(g)-(i)]. The $\varphi_{Zak}$ changes from the trivial ($\varphi_x$, $\varphi_y$) = (0,0) in Phase A to the nontrivial in ($\varphi_x$, $\varphi_y$) = (0, -π) Phase B, and the parity reversal at Y reveal their fundamentally different topological nature[56]. For the 2D-SSH-NM model, only cutting the nanoribbons along the periodic *x*-direction generates nontrivial topological edge states. It can be regarded as parallel 1D-SSH chains stacked along *y*-direction, similar to the 3D Te chins[18]. Unlike conventional strong topological properties, its topological phase originates from intra- and inter-unit modulatory transitions and generates direction-dependent topological edge states, which should be the weak topological properties[57-60].

**Realizations of 2D SSH model**

To realize the above 2D-SSH-NM model, we propose $C_4N_2$ and $C_6N_2Si_2$ monolayers as ideal candidates, where the structural parameters are summarized in Table S1. Phonon spectra confirm their dynamic stability in Figs. S4(a) and (b). The both monolayers feature degenerate Dirac nodal lines along XM paths and exhibit semimetallic character, as shown in Figs. 2(c)-(f). The detailed calculation method is shown in SM Part II. The flat band structures only emerge along XM rather than all high symmetry lines in the first BZ. Experimentally, the similar band structure has been detected in bulk $NbSi_{0.45}Te_2$, while quasi-1D Dirac nodal lines were theoretically predicted earlier in 2D carbon nitride monolayers[59,61]. The Kohn-Sham wavefunction and orbital-projected calculations near the nodal line confirm N-$p_z$ orbitals dominate nodal line formation [Fig. S5]. Then we only analyze N atoms in subsequent discussions. As shown in Figs. 2(a) and (b), both monolayers possess two unit-cell phases defined by alternating strong hopping (solid black lines) and weak hopping (dashed black lines). Although the two phases have the same band structures [Figs. 2(c)-(f)], the parity inversion at the Y indicates the possible topological phase transition. Cutting along weak hopping gives rise to trivial edge states, consistent with previous TB predictions [Fig. 2(g) and (i)]. Conversely, cutting along strong hopping generates nontrivial edge states, as shown in Figs. 2(h) and (j). The maximum bandgap is determined by the Y, where band inversion gives rise to these topologically protected edge states. Essentially, the 2D-SSH-NM model consists of 1D-SSH models in the *y*-direction arranged periodically along the *x*-direction, featuring strong anisotropy, as shown in Fig. 2(k). We further investigate uniaxial strains from −4% to 4% along *x*- and *y*-axis [Fig. S6]. The quasi-1D Dirac nodal lines remain robust under *x*/*y*-axis strain with only change the bandwidth of Dirac nodal lines along XM [Fig. S7(a)]. The bandwidth of Dirac nodal lines along XM is determined by hopping $t_3$, which is supported by TB results and has no effect to the topological properties, as shown in Fig. S7(c)[51]. Therefore, we realized the 2D-SSH-NM model in carbon-based monolayers.

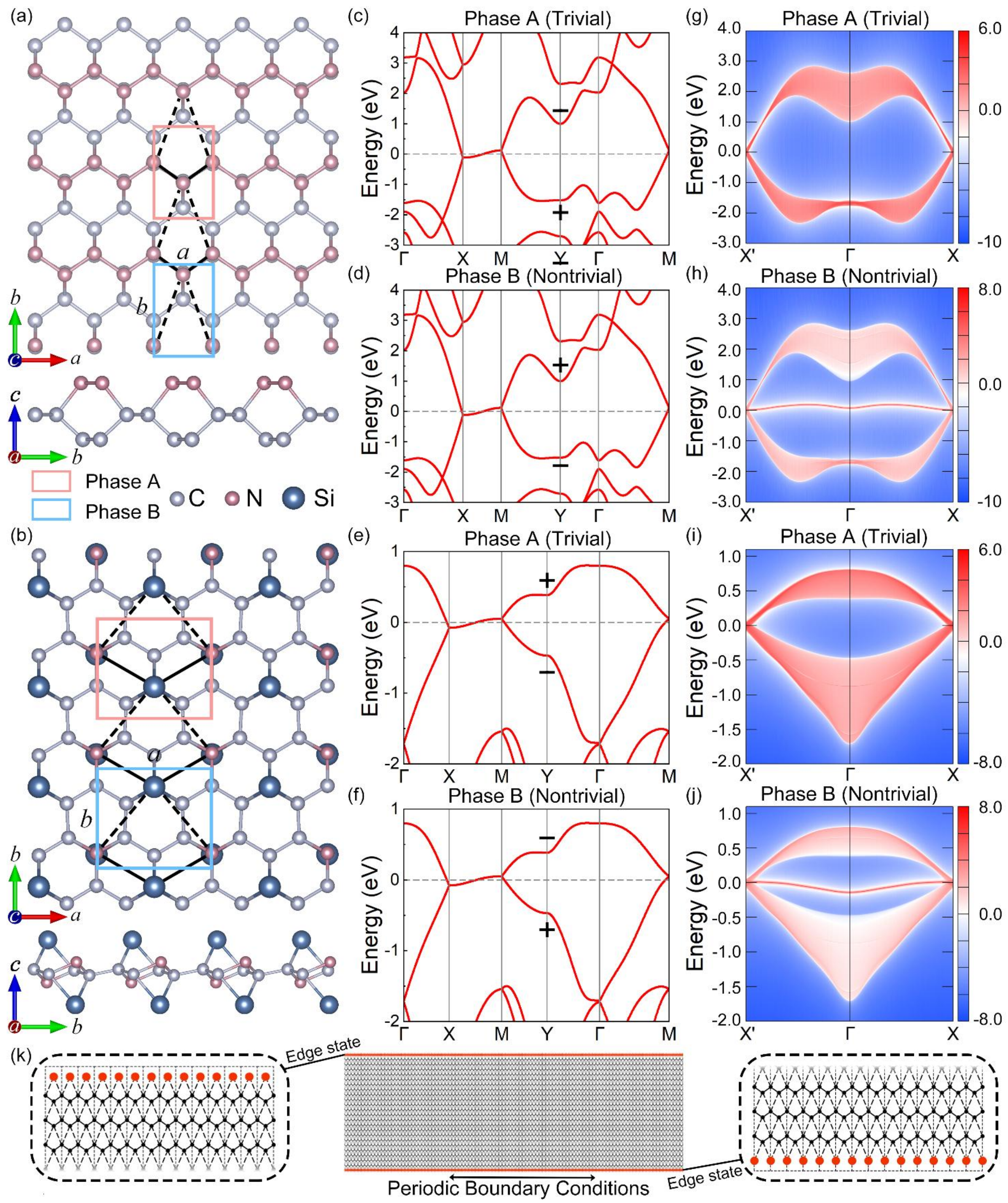

**FIG. 2.** (a)(b) Structures of $C_4N_2$ and $C_6N_2Si_2$ with two unit-cell phases (red for Phase A, blue for Phase B). (c)(d) Band structures of trivial/nontrivial $C_4N_2$ of Phase A/B, corresponding to the edge states in (g)(h). (e)(f) Band structures of trivial/nontrivial $C_6N_2Si_2$ of Phase A/B, corresponding to the edge states in (i)(j). (k) Spatial projection of edge states.

**Dirac-Mott AFM phase transition**

Further considering magnetism, the antiferromagnetism should be the ground state for $C_4N_2$ monolayer, where the two N atoms exhibit weak magnetic moment. The bandgap of the AFM state can be opened in the extended 2D-SSH model to break the limitations of prior nonmagnetic symmetric frameworks. We employ the TB method to calculate $\varphi_{\text{Zak}}$ and edge states features for both configurations [SM Part III]. The two phases have the same band structures [Fig. 1(f)]. The distinct $\varphi_{\text{Zak}}$ reveal their fundamentally different topological nature[50-53]. Phase A features ($\varphi_x, \varphi_y$) = (0,0) and trivial edge states [Fig. 1(i)], whereas Phase B has ($\varphi_x, \varphi_y$) = (0, -$\pi$) accompanied by nontrivial topological edge states, as shown in Fig. 1(l). The main difference between the 2D-SSH-NM and 2D-SSH-AFM model is whether there is a bandgap in 2D band structures and corresponding 1D edge states.

**Symmetry analysis for AM phase**

The 2D-SSH-AFM model corresponding to Fig. 1(c) can accurately reproduce the band structure of the $C_4N_2$ monolayer with collinear AFM state, but this model cannot match the real symmetry of the $C_4N_2$ monolayer. Since it belongs to the space group *Pma*2 (No. 28), we use two magnetic sublattices (A and B) and two nonmagnetic sites to fully describe it, as shown in Fig. 3(a). The fundamental real-space symmetry operations are $O = C_{2z}, G_x, M_x$. The $C_2$ is a spin-space 180° rotation that flips spin (↑↓), which is combined with real-space geometric operation $O$ as $[C_2||O]$. The $C_2$ is paired with time reversal $T$ in $[C_2||T]$. It can be found that the product of $[C_2||C_{2z}]$ and the spin-only symmetry $[C_2||T]$ equals $[E||TC_{2z}]$. This means $[E||TC_{2z}]$ $E$ (↑, $\boldsymbol{k}$) = $E$ (↓, $\boldsymbol{k}$), corresponding to the conventional AFM state[46]. To obtain the nontrivial AFM state with spin-splitting (AM state), three strategies can be chosen. (i) Breaking the $T$ and preserving the $C_{2z}$ (2D-SSH-CPL model). It can be realized by the Floquet engineering of CPL[48]. (ii) Breaking the $C_{2z}$ and preserving the $T$ (2D-SSH-RAD model). It can be realized by the RAD along $x$-axis between the two magnetic atoms and the two nonmagnetic atoms[39], as shown in Fig. 3(a). (iii) Breaking the $C_{2z}$ and breaking the $T$ at the same time (2D-SSH-RAD-CPL model). After (ii), applying CPL can realize it. In the following, we will discuss the three models by TB methods.

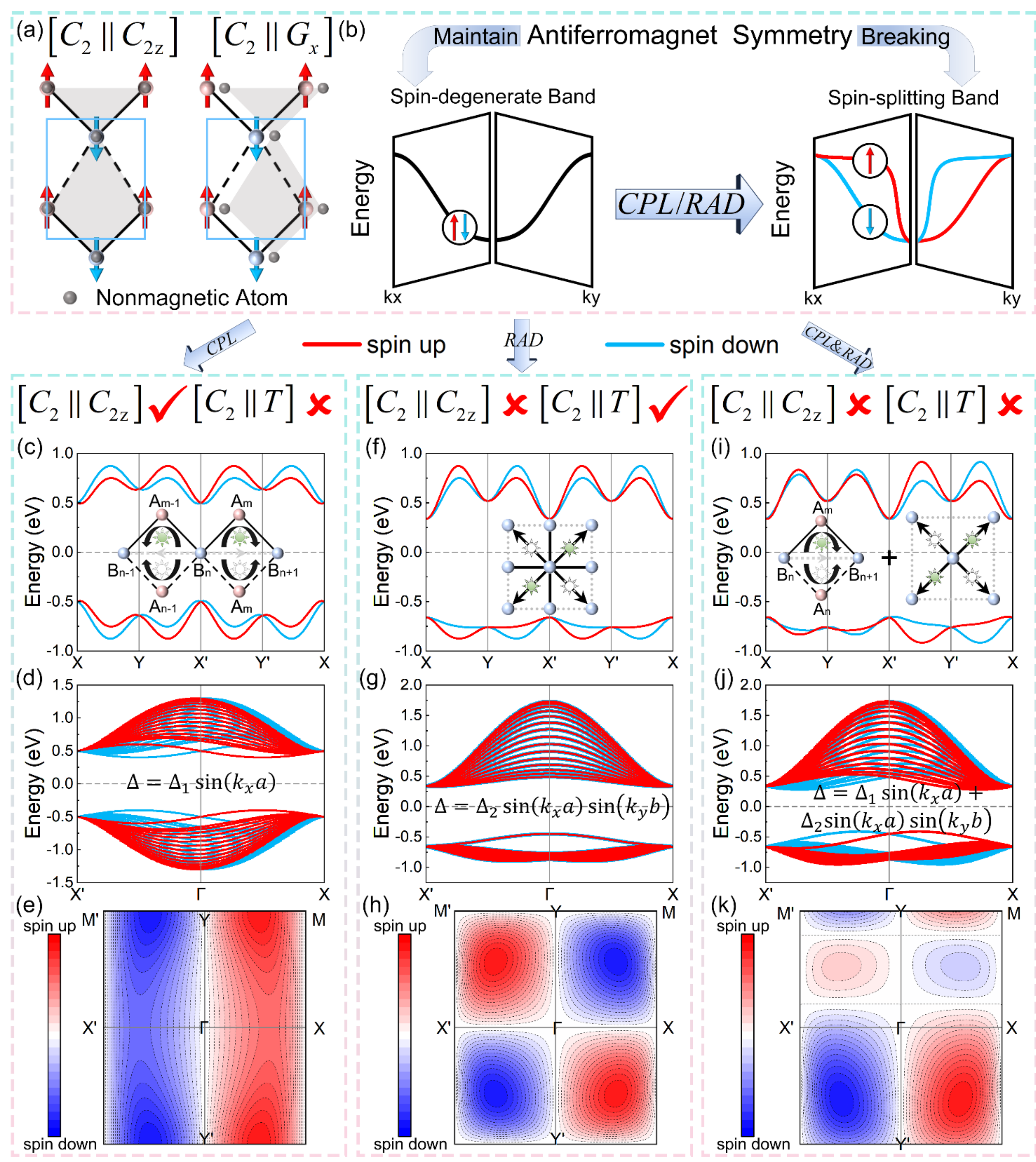

**FIG. 3.** (a) Two extended SSH models. (b) Band structures of AFM and AM states. (c)(d)(e) Band structure, topological edge states, and odd-parity *p*-wave AM order under CPL. (f)(g)(h) Band structure, topological edge states, and even-parity *d*-wave AM order under RAD. (i)(j)(k) Band structure, topological edge states, and mixed-parity *p*-wave AM order under RAD-CPL.

**TB method**

(i) 2D-SSH-CPL model. The right-handed CPL generates effective Haldane terms by wo optical transition pathways[48], as shown in Fig. 3(c). From the TB methods, the effective Haldane term induced by CPL can produce two hopping terms along $x$-axis with opposite symbols and unequal values ($B_n \rightarrow A_n \rightarrow B_{n+1}$ and $B_n \rightarrow A_m \rightarrow B_{n+1}$) due to $|t_1| \neq |t_2|$, leading the form $\Delta = \Delta_1 \sin(k_x a)$ [SM Part IV]. In this case, the Haldane term induced by CPL breaks $[C_2||T]$ while preserving $[C_2||C_{2z}]$. As shown in Fig. 3(c), the band structure under CPL modulation has spin-splitting and satisfy $E(\uparrow, \boldsymbol{k}) = E(\downarrow, -\boldsymbol{k})$, showing the AM signatures of with odd-parity $p$-wave[62]. Upon switching from right-handed to left-handed CPL, the spin-splitting direction is reversed[46]. As shown in Fig. 3(d), the corresponding 1D edge states can also show the AM signatures with odd-parity p-wave, where all non-boundary sites still satisfy the condition with opposite signs and unequal values. For boundary A (B) sites, only one hopping term along the $x$-axis survives, where the hopping path associated with $|t_2|$ vanishes, whereas the path with $|t_1|$ remains [Fig. S8]. We therefore conclude that the 1D-SSH model [Fig. 1(a)] in the AFM phase under CPL can also host AM properties. Nevertheless, both 1D-SSH models with $|t_1| \neq |t_2|$ and $|t_1| = |t_2|$ can realize the AM state, implying no coupling between the Floquet-engineered AM and SSH topological states. By contrast, $|t_1| \neq |t_2|$ is a prerequisite for Floquet-engineered AM coexisting with SSH weak topological states in the 2D-SSH model, which points to a coupled underlying physical mechanism.

(ii) 2D-SSH-RAD model. The RAD along $x$-axis breaks $[C_2||C_{2z}]$ and preserves $[C_2||T]$. The band structure is dominated by inequality of hopping terms (A↔A and B↔B) along diagonal directions due to $|t_1| \neq |t_2|$, which are described by $\Delta = \Delta_2 \sin(k_x a)\sin(k_y b)$ [SM Part V]. The AM band structure is spin-splitting [Fig. 3(f)], satisfying $E(\uparrow, \boldsymbol{k}) = E(\uparrow, -\boldsymbol{k})$ and $E(\downarrow, \boldsymbol{k}) = E(\downarrow, -\boldsymbol{k})$, showing the AM signatures of even-parity $d$-wave[62]. Upon switching from positive direction of the $x$-axis to negative direction of the $x$-axis, the spin-splitting direction is reversed [Fig. S8]. Different from the 2D-SSH-CPL model, the corresponding topological edge states carry no altermagnetic signatures due to the 2D system of even-parity $d$-wave [Fig. 3(g)]. (iii) 2D-SSH-RAD-CPL model. Starting from the (ii), additional CPL can further break $[C_2||T]$ and introduces a combined modulation term $\Delta = \Delta_1\sin(k_x a) + \Delta_2\sin(k_x a)\sin(k_y b)$, becoming a $p$-wave AM phase from the $d$-wave AM phase [Fig. 3(i)]. Increasing CPL intensity evolves the hybrid phase into complete $p$-wave AM phase similar to the 2D-SSH-CPL model [Fig. S10]. The CPL also activates AM signatures of even-parity $d$-wave in the edge states [Fig. 3(j)]. However, this AM phase is a mixed-parity rather than odd-/even-parity, because $E(\uparrow, \boldsymbol{k}) \neq E(\uparrow, -\boldsymbol{k})$ $[E(\downarrow, \boldsymbol{k}) \neq E(\downarrow, -\boldsymbol{k})]$ and $E(\uparrow, \boldsymbol{k}) = E(\downarrow, -\boldsymbol{k})$. This mixed-parity AM phase originates from the disappearance of the real-space structural symmetry. The system only preserves the symmetry of the Floquet-lattice interaction, which is equivalent to $[C_2||M_x]$.

**DFT calculations**

To validate the above SSH-AM model in real materials, we screen 2D magnetic candidates meeting the symmetry requirements. The $C_8N_2$ monolayer host the SSH weak topological properties with the AFM state (magnetic moments localized at raised carbon atoms)[59]. The detailed parameters are listed in Table S1 and phonon spectrum show entirely positive frequencies [Fig. S4(c)]. The space group $Pma2$ (No. 28) is in good agreement with the 2D-SSH-AFM model, resulting in spin-degenerate band structure [Fig. 4(c)], and the corresponding edge states are nontrivial [Fig. 4(g)]. Appling the CPL, the AM state of odd-parity $p$-wave can be found in the band structure and edge states, corresponding to the 2D-SSH-CPL model, and the maximum spin-splitting gap reaches 135 meV [Fig. 4(d) and (h)]. Applying left-handed CPL, the energy positions of the spin-up and spin-down states interchange at the same $\boldsymbol{k}$ point [Fig. S11(a) and (b)]. Besides the $C_8N_2$, $C_{12}H_6$ monolayer is another viable 2D material to realize our topological magnetic modulation. Its crystal structure shown in Fig. 4(b), and full parameters are provided in Table S1. Phonon spectrum of $C_{12}H_6$ shows positive frequencies [Fig. S4(d)]. The monolayer belongs to the space group *Pc* (No. 7) with preserving symmetry $G_x$. Since only the $[C_2||G_x]$ is preserved without $[C_2||C_{2z}]$, the band structure is intrinsic spin-splitting [Fig. 4(e) and 4(i)], which is an AM signature of even-parity *d*-wave and the maximum spin-splitting gap reaches 61 meV. Constrained by the *d*-wave symmetry, the system cannot support observable AM edge states [Fig. S12(b)], corresponding to the 2D-SSH-RAD model. In fact, the positive RAD of the hydrogen atom corresponds to the antiferroelectric property. For another antiferroelectric structure (negative RAD), the energy positions of the spin-up and spin-down states interchange at the same $\boldsymbol{k}$ point [Fig. S11(c) and (d)]. Further applying the CPL can further break $T$, a mixed altermagnetic phase can be produced [Fig. 4(f)]. It removes the AM order of even-parity *d*-wave for 2D system and induces AM edge states [Fig. S12(c) and (d)], corresponding the 2D-SSH-RAD-CPL model. Therefore, the AM phase of mixed-parity *p*-wave can be realized in the $C_{12}H_6$ monolayer with CPL. At $eA/\hbar=0.3\ \AA^{-1}$, the maximum spin-splitting value reaches 42 meV, revealing effective Floquet engineering modulation.

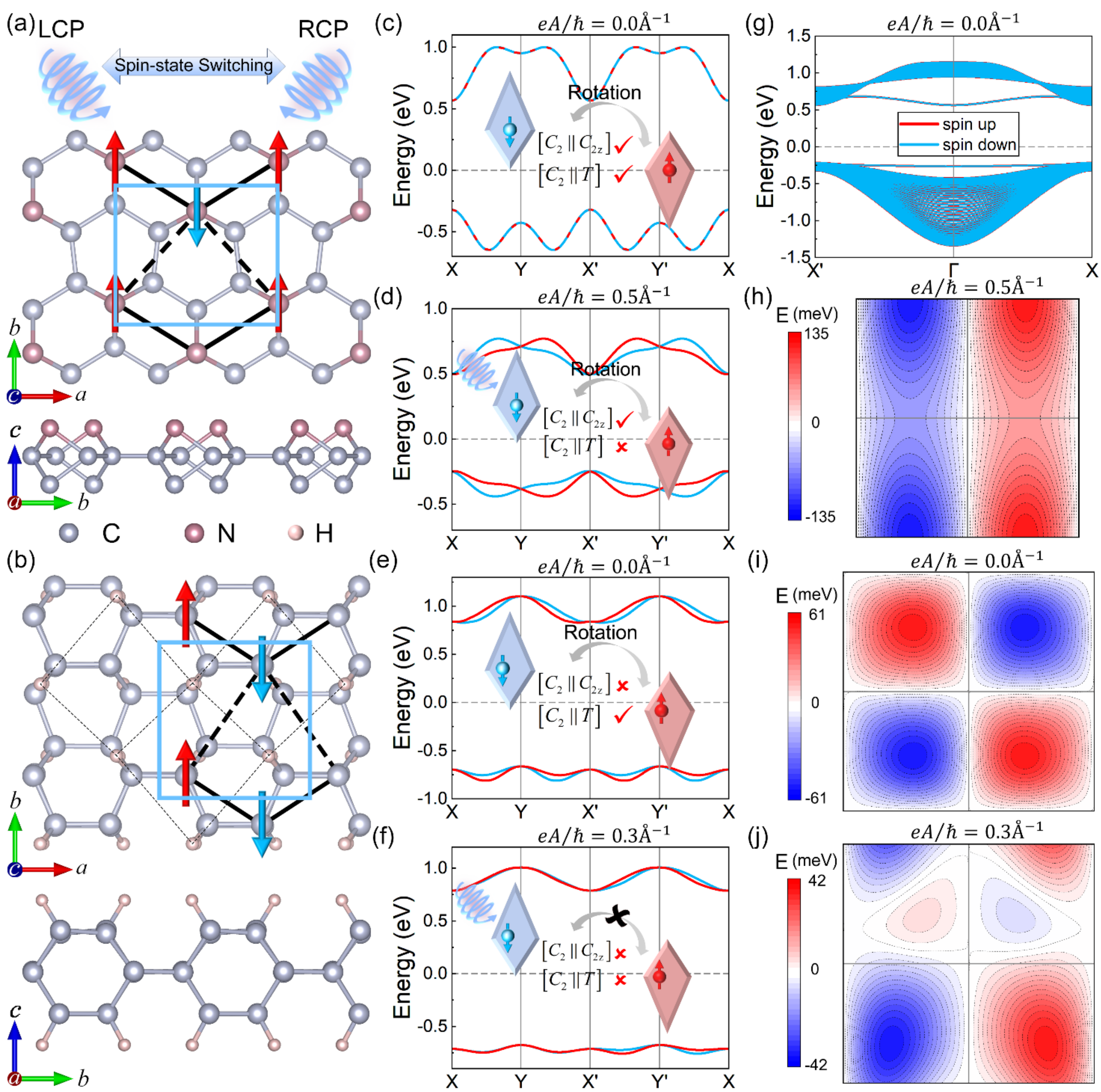

**FIG. 4.** (a)(b) Structures of $C_8N_2$ and $C_{12}H_6$. (c)(d) Band structures of $C_8N_2$ without and with right-handed CPL. (e)(f) Band structures of intrinsic and CPL-modulated $C_{12}H_6$. (g) Edge states of spin-degenerate $C_8N_2$. (h) AM phase of odd-parity *p*-wave in CPL-modulated $C_8N_2$. (i) AM phase of even-parity *d*-wave-type in intrinsic $C_{12}H_6$. (j) Mixed-parity AM phase in CPL-modulated $C_{12}H_6$.

**Conclusion**

This work constructs five 2D-SSH TB models to describe the coupled phases of SSH weak topological state and AM order. For the NM state, the anisotropic topological state can be achieved by changing the selection of unit cell, which should be the SSH weak topological properties with edge states determined by Zak phase due to the $|t_1| \neq |t_2|$. The results from the 2D-SSH-NM of TB models are in good agreement with those from the DFT calculations. For the conventional AFM state, the collinear antiferromagnetism could be achieved, where the spin-degenerate band structure can be achieved in the 2D/1D system, corresponding to the 2D-SSH-AFM model. The topological results should be preserved due to the $|t_1| \neq |t_2|$. For the AM state, the Floquet engineering introduces the AM phase of odd-parity *p*-wave (2D-SSH-AM-CPL model) while the RAD introduces the AM phase of even-parity *d*-wave (2D-SSH-AM-RAD model). From the TB models, we show that the origin of the AM phase is also the inequality between $|t_1|$ and $|t_2|$. Furthermore, by combining the Floquet engineering and the RAD, mixed-parity AM phase could be realized, which is in good agreement with the 2D-SSH-AM-RAD-CPL model. In summary, the realizations of topological state and AM order depend on the inequality between $|t_1|$ and $|t_2|$, and our proposed 2D-SSH models provide an excellent platform for coupling of the two nontrivial electronic structures.

**ACKNOWLEDGMENT**

This work is supported by the Natural Science Foundation of Hebei Province (Grant No. A2025202032).

**DATA AVAILABILITY**

The data that support the findings of this article are not publicly available upon publication because it is not technically feasible and/or the cost of preparing, depositing, and hosting the data would be prohibitive within the terms of this research project. The data are available from the authors upon reasonable request.